\documentclass[10pt,twocolumn,letterpaper]{article}

\usepackage[pagenumbers]{wacv} 

\usepackage{graphicx}
\usepackage{amsmath}
\usepackage{amssymb}
\usepackage{booktabs}
\usepackage{comment}
\usepackage{caption}
\usepackage{subcaption}
\usepackage{overpic}
\usepackage{rotating}
\usepackage{subeqnarray}
\usepackage{array,multirow}
\usepackage{tabularx}
\usepackage{enumerate}
\usepackage{color}
\usepackage[dvipsnames, table]{xcolor}
\usepackage{colortbl}
\definecolor{wacvgreen}{rgb}{0.21,0.74, 0.39}
\usepackage[pagebackref,breaklinks,colorlinks,citecolor=wacvgreen]{hyperref}
\usepackage[accsupp]{axessibility}
\usepackage[capitalize]{cleveref}
\crefname{section}{Sec.}{Secs.}
\Crefname{section}{Section}{Sections}
\Crefname{table}{Table}{Tables}
\crefname{table}{Tab.}{Tabs.}
\newcommand{\RNum}[1]{\uppercase\expandafter{\romannumeral #1\relax}}

\newcommand{\eref}[1]{Equation~\ref{#1}}

\newcommand{\disclaimer}{%
    \vspace{-1.5cm}
    \begin{center}
        \normalsize\textit{To appear in Proceedings of the \emph{IEEE/CVF Winter Conference on Applications of Computer Vision (WACV) 2024.}}        
    \end{center}
    \vspace{1.0cm}
}
\definecolor{Gray}{gray}{0.95}
\definecolor{LightCyan}{rgb}{0.88,1,1}
\definecolor{Yellow}{rgb}{1,0.9,0.7}
\definecolor{Red}{rgb}{1,0.8,0.8}
\definecolor{Green}{rgb}{0.7,1,0.7}
\definecolor{Blue}{rgb}{0.8,1,1}

\hypersetup{
     pdftitle={BSRAW: Improving Blind RAW Image Super-Resolution},
     pdfsubject={RAW Super-Resolution},
     pdfauthor={Marcos V. Conde, Florin Vasluianu, Radu Timofte}
}

\def\wacvPaperID{} 
\def\confName{WACV}
\def\confYear{2024}

\begin{document}


\title{\disclaimer BSRAW: Improving Blind RAW Image Super-Resolution}

\author{Marcos V. Conde, Florin Vasluianu, Radu Timofte\\
{\normalsize Computer Vision Lab, CAIDAS \& IFI, University of Würzburg}\\
{\tt\small \{marcos.conde, florin-alexandru.vasluianu\}@uni-wuerzburg.de}\\
{\tt\normalsize {\url{https://github.com/mv-lab/AISP}}}
}

\maketitle


\begin{abstract}
In smartphones and compact cameras, the Image Signal Processor (ISP) transforms the RAW sensor image into a human-readable sRGB image. Most popular super-resolution methods depart from a sRGB image and upscale it further, improving its quality. However, modeling the degradations in the sRGB domain is complicated because of the non-linear ISP transformations. Despite this known issue, only a few methods work directly with RAW images and tackle real-world sensor degradations.

We tackle blind image super-resolution in the RAW domain. We design a realistic degradation pipeline tailored specifically for training models with raw sensor data. Our approach considers sensor noise, defocus, exposure, and other common issues. Our BSRAW models trained with our pipeline can upscale real-scene RAW images and improve their quality. As part of this effort, we also present a new DSLM dataset and benchmark for this task.
\end{abstract}


\section{Introduction}
\label{sec:introduction}


The specialized on-board Image Signal Processor (ISP) transforms RAW sensor readings into more refined and perceptually meaningful RGB representations~\cite{conde2022modelbased,heide2014flexisp,schwartz2018deepisp,zhang2019zoom}.
Photographers often opt for RAW over RGB processing to yield perceptually superior images, attributing this preference to two key benefits of RAW data:
(i) RAW information carries a broader data range, generally ranging from 12-14 bits, compared to the 8-bit RGB output from the ISP.
(ii) RAW data linearly correlates with scene radiance, which simplifies the correction of degradations such as noise.
Conversely, ISP operations are non-linear and undergo an irrevocable loss of information across various stages~\cite{brooks2019unprocessing, karaimer2016software}, complicating subsequent image restoration~\cite{brooks2019unprocessing, xu2019rawsr}.
For these considerations, RAW imagery is a more advantageous starting point than RGB for a multitude of low-level vision applications like image denoising~\cite{abdelhamed2018high, zamir2020cycleisp, brooks2019unprocessing, mildenhall2018burst, yue2020supervised, ma2022elmformer}, color balance~\cite{Hernandez-Juarez_2020_CVPR}, deblurring~\cite{liang2020raw}, exposure adjustment~\cite{zhang2022low, hasinoff2016burst, chang2021low}, and image super-resolution~\cite{xu2019rawsr, zhang2019zoom, qian2019trinity, yue2022real, conde2022swin2sr}.


Nevertheless, only a small fraction of low-vision research directly engages with RAW data, mainly due to the greater abundance and accessibility of general-purpose sRGB images. 
Consequently, the most popular deep learning architectures for image restoration predominantly operate on RGB images~\cite{zamir2020cycleisp, zamir2022restormer, wang2021uformer, chen2022simple}. The majority of cutting-edge Single Image Super-Resolution (SISR) techniques~\cite{liang2021swinir, ji2020realsr, agustsson2017ntire, zhang2020usrnet, dong2014srcnn} rely on deep convolutional networks or Transformer~\cite{vaswani2017transformer} architectures, and operate in the RGB color space.
Yet, these approaches have notable drawbacks. Primarily, they are often trained on artificially generated data, leading to poor generalization on real-world scenes. Secondly, modeling accurately the degradations in RGBs is challenging due to the non-linear transformations and information loss incurred during ISP processing (RAW to RGB).

The classical single-image super-resolution model~\cite{elad1997restoration, kai2021bsrgan, xu2019rawsr} is formulated as:

\begin{equation}\label{eq:sisr_degradation}
  \mathbf{y}\!= \!(\mathbf{x}\otimes \mathbf{k})\!\downarrow_{\bf{s}}\! + \,\mathbf{n}
\end{equation}

It assumes the observed -or captured- low resolution (LR) image $\mathbf{y}$ is obtained from an underlying high resolution (HR) image $\mathbf{x}$ by applying a degradation kernel (or PSF) $\mathbf{k}$~\cite{efrat2013accurate}, followed by a downsampling operation $\downarrow_{\bf{s}}$ with scale factor $\bf{s}$ (\eg Bicubic~\cite{yang2010image,Timofte2014AAA}), and the addition of the noise $\mathbf{n}$~\cite{hasinoff2014photon}. 
The more realistic these factors are modelled and applied, the better the restoration and SISR models perform and generalize in real-world scenarios~\cite{xu2019rawsr, kai2021bsrgan}.

\begin{figure}[!ht]
     \centering
     \begin{subfigure}{\linewidth}
         \centering
         \includegraphics[width=\linewidth]{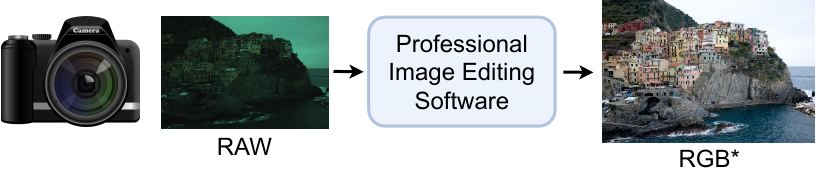}
         \caption{Professional photography image editing~\cite{fivek, karaimer2016software}. The software (\eg Adobe Lightroom) usually applies restoration on the RAW image, and further ISP and enhancement steps.
         \vspace{2mm}}
         \label{fig:main-a}
     \end{subfigure}
     \begin{subfigure}{\linewidth}
         \centering
         \includegraphics[width=\linewidth]{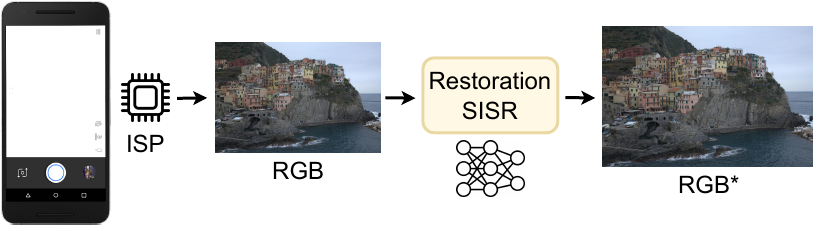}
         \caption{Image Restoration and Super-Resolution in the RGB domain as offline post-processing~\cite{zamir2022restormer, kai2021bsrgan, liang2021swinir, chen2022simple, wang2021uformer} --- most popular.
         \vspace{2mm}}
         \label{fig:main-b}
     \end{subfigure}
     \begin{subfigure}[b]{\linewidth}
         \centering
         \includegraphics[width=\linewidth]{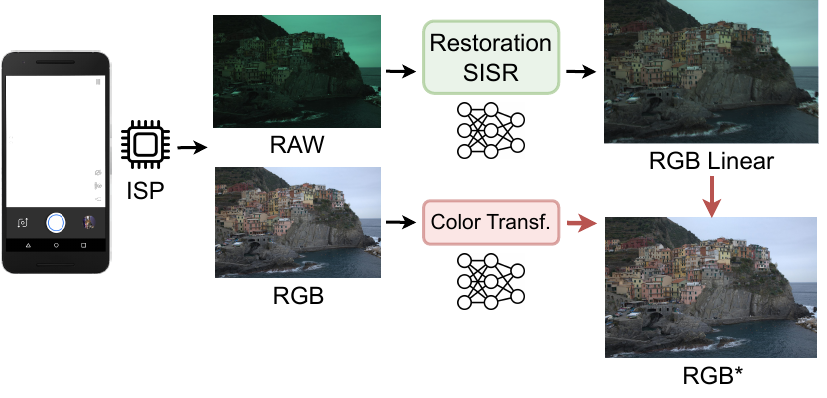}
         \caption{Xu~\etal~\cite{xu2019rawsr, xu2020exploiting} approach leverages RAW and RGB images for real scene super-resolution using a double-branch network.
         \vspace{2mm}}
         \label{fig:main-c}
     \end{subfigure}
     \begin{subfigure}[b]{\linewidth}
         \centering
         \includegraphics[width=\linewidth]{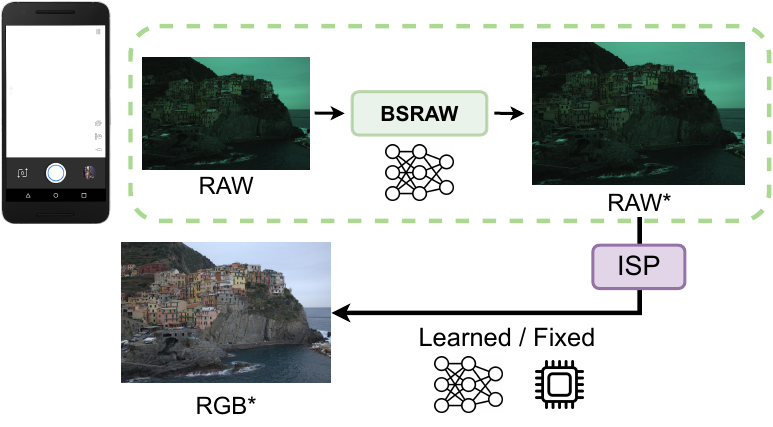}
         \caption{\textbf{Our approach BSRAW}. We focus on \textbf{B}lind \textbf{S}uper-\textbf{R}esolution of \textbf{RAW} images. Our model operates before the ISP as a plug \& play block, not as post-processing \emph{(b,c)}, thus it can benefit other downstream tasks.}
         \label{fig:main-d}
     \end{subfigure}
     \caption{Comparison of image restoration and SISR pipelines. Note that previous approaches (b) and (c) depend on the in-camera ISP, meanwhile our BSRAW method can complement any (learned) ISP, as a pre-processing block. We refer as RAW* and RGB* to the enhanced images. Sample image from the Adobe MIT5K dataset~\cite{fivek}.
     }
     \label{fig:main-wild}
\end{figure}

This introduces the need for more complex degradation pipelines, to reduce the generalization gap observed in the real world domain. Recently, Zhang~\etal~\cite{kai2021bsrgan} designed a practical degradation pipeline for the RGB domain SISR, achieving promising results in real scene super-resolution.

RAWSR from Xu~\etal~\cite{xu2019rawsr, xu2020exploiting} is one of the first works that apply a simple degradation pipeline (Eq.~\ref{eq:sisr_degradation}) for synthesizing low-resolution RAW images and train SISR models that successfully generalize on real scenes. Their approach is illustrated in Figure~\ref{fig:main-c}, ultimately they aim at enhancing RGB images by leveraging both the sensor RAW data and the RGB produced by the ISP. For this reason, their studies are performed on the resultant RGBs, and their method can be seen as an offline \emph{post-processing (always after the ISP)} for image enhancement and photo-finishing. This is similar to most popular image restoration and SISR techniques~\cite{zamir2022restormer, kai2021bsrgan, liang2021swinir, chen2022simple} -see Figure~\ref{fig:main-b}-.\\

In this work we study \textbf{RAW SISR} -see Figure \ref{fig:main-d}-, using exclusively RAW sensor data, thus, the signal depends only on the scene and the imaging system (\eg light sources, user-camera interaction). Moreover, the degradations such as noise~\cite{abdelhamed2018high, brooks2019unprocessing} are not altered by the ISP stages~\cite{delbracio2021mobile, karaimer2016software}. We build a  realistic RAW degradation pipeline that extends and improves previous methods from Zhang~\etal~\cite{kai2021bsrgan} and Xu~\etal~\cite{xu2019rawsr, xu2020exploiting}. Our baseline model, BSRAW, is an efficient blind RAW SISR method, trained using our degradation pipeline, and real data from DSLR and DSLM cameras. The \emph{main contributions} of this paper are:

\begin{itemize}
  \item[1)] A controllable degradation pipeline to train deep neural models using RAW images and considering realistic downsampling, noise and blur.
  \item[2)] A new dataset for RAW SISR including digital single-lens mirrorless (DSLM) cameras.
  \item[3)] We study BSRAW and other methods for this task, and we provide a new benchmark for RAW SISR.
\end{itemize}

\section{Related Work}
\label{sec:related_work}

\paragraph{Image Signal Processing}

Digital camera systems incorporate a specialized in-camera Image Signal Processor (ISP)~\cite{karaimer2016software, delbracio2021mobile, heide2014flexisp, xing2021invertible, schwartz2018deepisp}, which converts the captured RAW sensor data into images that are both less deteriorated and visually appealing to human perception.

In the recent years many deep learning approaches have been proposed to learn the ISP transformation~\cite{chen2018learning, liu2022deep}. Some popular methods in this direction are CycleISP~\cite{zamir2020cycleisp}, CameraNET~\cite{liang2021cameranet},  and PyNET~\cite{ignatov2020replacing}.

In this work, we use ISP pipelines~\cite{karaimer2016software} with known parameters and operations, however, our proposed method can help to improve learned ISPs for real scene processing.

\paragraph{Degradation Models}

Creating authentic degradation models for synthesizing data is a critical endeavor in the field of low-level vision. Standard degradation models frequently adopt a sequential approach involving blur, downsampling, and noise injection (as described in Eq.~\ref{eq:sisr_degradation}). Since most degradation models are far from the real-world complexity, some works aimed at solving such problem by designing complex degradation pipelines~\cite{lugmayr2019unsupervised, bulat2018learn, kai2021bsrgan, xu2019rawsr}.

Zhang~\etal~\cite{kai2021bsrgan, zhang2022practical} devised a pragmatic degradation methodology specifically tailored for training sRGB image restoration models, achieving encouraging outcomes in Single Image Super-Resolution (SISR). Extensions of this approach to video denoising have also been reported~\cite{cao2022practical}.

Xu~\etal~\cite{xu2019rawsr, xu2020exploiting} implement the degradation equation (Eq.~\ref{eq:sisr_degradation}) to synthesize low-resolution RAW images and then train models that exhibit strong generalization capabilities in real-world scenarios. In their work, they use heteroscedastic Gaussian noise, varying disk kernels for defocus blur, and minimal motion blur.

Little research exists on elaborating more intricate degradation phenomena, apart from noise, in the RAW image space~\cite{hasinoff2014photon, liang2020raw, abdelhamed2018high}. In response, we introduce improvements such as enhanced noise profiles, anisotropic defocusing, motion blur variation, exposure inconsistencies, and compression artifacts. Details of our enhanced degradation model are elaborated in Section~\ref{sec:method}.

\subsection{Image Super-Resolution and Restoration}
\label{ssc:flexible_SISR}

In the recent years blind image restoration and super-resolution using deep learning has become a popular task~\cite{dong2014srcnn, zhang2018residualdense, wang2018esrgan, liang2021swinir, kai2021bsrgan, chen2022simple, zamfir2023towards,conde2023perceptual}, this is because blind methods do not require prior knowledge about the degradation factors or the camera sensor, therefore they are general purpose models \ie practically, such methods can process any degraded RGB image. SwinIR~\cite{liang2021swinir} is an example of general purpose SISR and restoration method. 

The vast majority of blind restoration and SISR methods operate in the RGB color space. We also focus on the blind super-resolution problem, yet, in the RAW domain.

There are far fewer works that solve complex inverse problems in the RAW domain.
Most methods focus on RAW image denoising, since estimating and removing the noise on linear raw data is well-studied~\cite{abdelhamed2018high, brooks2019unprocessing, mildenhall2018burst, chen2018learning}. SIDD~\cite{abdelhamed2018high} and DND~\cite{plotz2017benchmarking} are the most popular RAW denoising benchmarks. These benchmarks show that \emph{state-of-the-art} methods can achieve high reconstruction ($>50$dB PSNR) given a noisy RAW image, indicating that new challenging data should be used. Liang~\etal studied RAW image deblurring on a simple setup using images from one DSLR camera~\cite{liang2020raw}. We also find works focused on exposure correction for low-light image enhancement~\cite{hasinoff2016burst, chang2021low, huang2022towards, chen2018learning}. Xu~\etal introduced RAWSR~\cite{xu2019rawsr, xu2020exploiting} for super-resolution and enhancement using both RAW and RGB images as inputs.
%
\vspace{1mm}

%
We aim to study the RAW SISR problem assuming (i) unknown real-world degradations, and (ii) cross-device generalization \ie our method can be applied to images from different sensors, as the most popular RGB restoration models~\cite{zamir2022restormer, liang2021swinir,chen2022simple}.

\begin{figure*}[ht]
    \centering
    \setlength{\tabcolsep}{0pt}
    \begin{tabular}{c c c c}
    \includegraphics[width=0.24\linewidth]{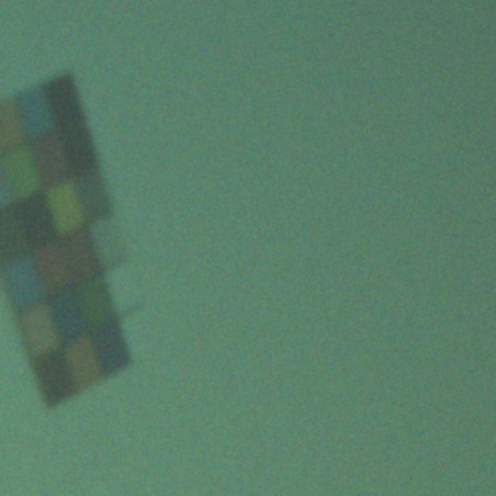} & 
    \includegraphics[width=0.24\linewidth]{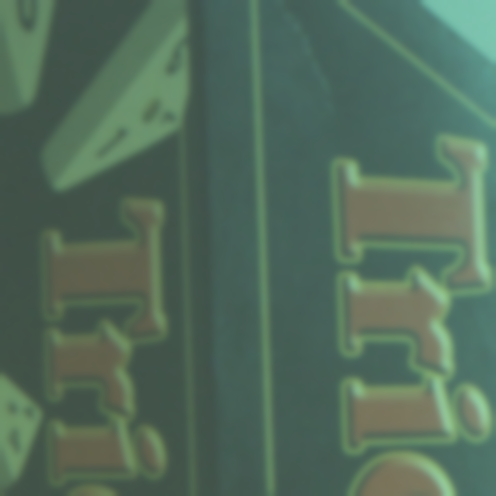} &
    \includegraphics[width=0.24\linewidth]{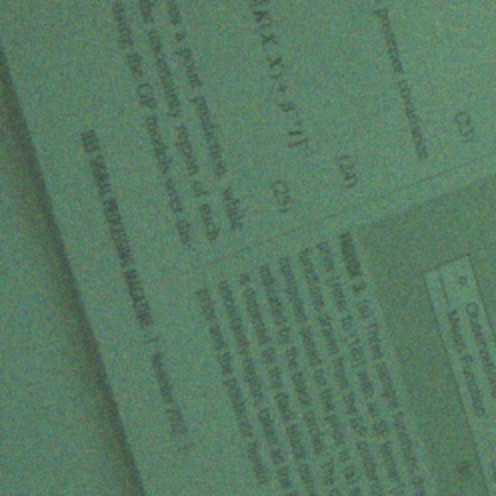} &
    \includegraphics[width=0.24\linewidth]{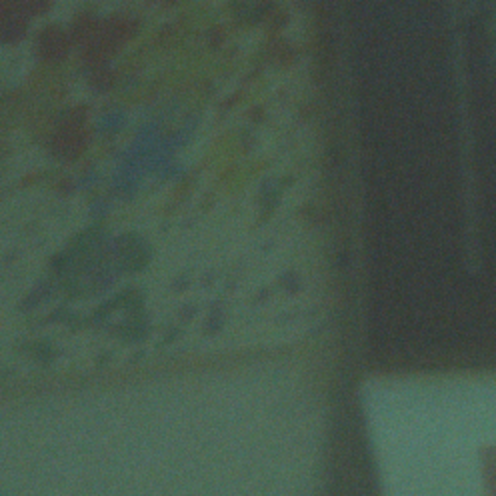} \\
    \includegraphics[width=0.24\linewidth]{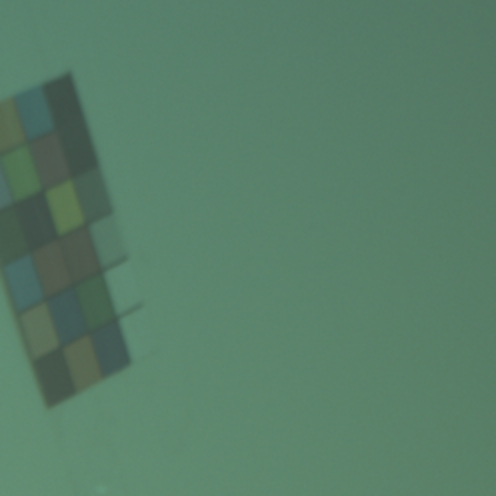} & 
    \includegraphics[width=0.24\linewidth]{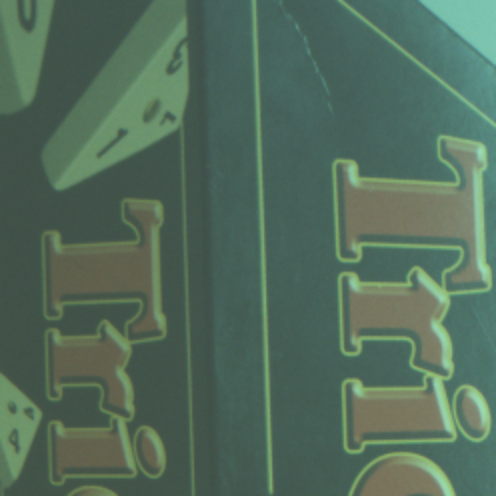} &
    \includegraphics[width=0.24\linewidth]{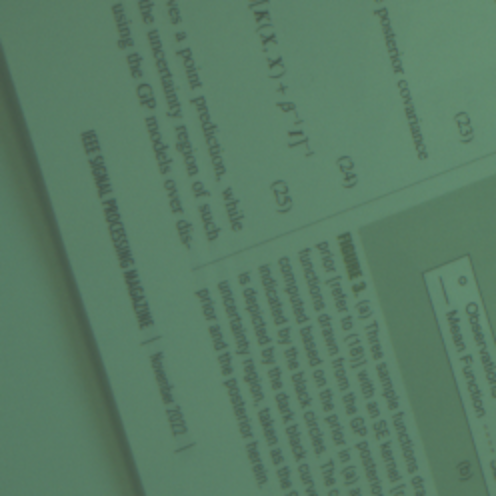} &
    \includegraphics[width=0.24\linewidth]{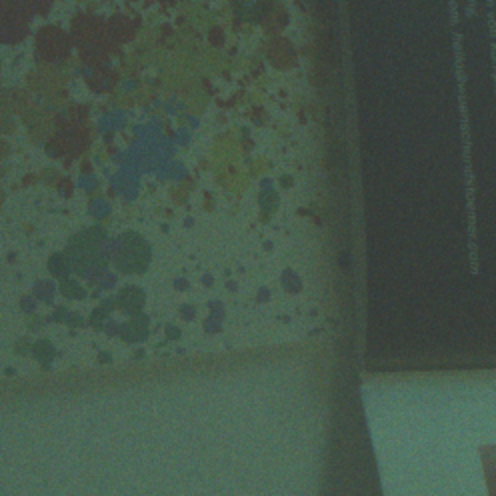} \\
    \end{tabular}
    \caption{Different variants of our dataset \textbf{test images}, after applying our degradation model. (Up) Degraded images, (Bot.) Clean images. The RAW images were acquired from multiple sensors (\eg Nikon D700 and NIkon Z7). RAW images are visualized after demosaicing, gamma correction and tone mapping. Best viewed in electronic version.}
    \label{fig:deg_samples}
\end{figure*}

\section{Controllable RAW Degradation Pipeline}
\label{sec:method}

We improve previous degradation models to synthesize realistic degraded RAW images~\cite{xu2019rawsr} for training SR models. We consider the most common limiting factors related to the acquisition and the processing of high quality RAW images: (i) noise and low resolution -- related to the size of the sensor~\cite{karaimer2016software}. (ii) exposure. (iii) motion and defocus blur -- related to stabilization issues. Note that previous methods~\cite{xu2019rawsr, xu2020exploiting} did not consider multiple noise profiles, or complex blur and PSFs. We provide samples from our dataset in Figure~\ref{fig:deg_samples}.

\subsection{Noise}
\label{section:noise}

Noise ($\mathbf{n}$) is an omnipresent element across all stages of image capture and processing~\cite{abdelhamed2018high, yoshimura2023rawgment}. Its occurrence is influenced by a multitude of variables, such as exposure duration and lighting conditions. In the domain of photography, noise removal —or denoising— commonly takes place in the RAW stage, where its linear characteristics have been thoroughly investigated~\cite{brooks2019unprocessing, hasinoff2014photon, hasinoff2016burst, abdelhamed2018high}. This makes it more manageable to address prior to the non-linear transformations introduced by the ISP~\cite{karaimer2016software}.
Most methods use Homoscedastic Gaussian noise~\cite{kai2021bsrgan, kai2017dncnn} in their degradation models. Xu~\etal~\cite{xu2019rawsr} use sharper Heteroscedastic Gaussian distribution~\cite{plotz2017benchmarking, abdelhamed2018high}.

We adopt a more practical shot-read noise~\cite{hasinoff2014photon, brooks2019unprocessing, zhang2021rethinking}. In \eref{eq:nat_noise}, we can observe the intensity $y$ as a sample of a Gaussian distribution having as mean the input signal $x$ and the variance depending on the $\lambda_{s}$ (shot) and $\lambda_{r}$ (read) parameters~\cite{brooks2019unprocessing}. This is derived from a Poisson-Gaussian noise model~\cite{wang2020practical}:

\begin{equation}
y \sim \mathcal{N}(\mu=x, \sigma^2=\lambda_{r} + \lambda_{s}x)
\label{eq:nat_noise}
\end{equation}

Utilizing Zhang \etal methodology~\cite{zhang2021rethinking}, we derive noise profiles specific to our imaging devices, in addition to adopting established profiles from SIDD~\cite{abdelhamed2018high, Abdelhamed_2019_CVPR_Workshops} and DND~\cite{plotz2017benchmarking} for smartphone and DSLR sensors, respectively. This diverges from earlier works that relied on a singular noise distribution~\cite{xu2019rawsr, brooks2019unprocessing}. Our approach incorporates \textbf{varied real-world noise profiles}, which manifest stochastically. We also emulate the interaction effects (e.g., attenuation, amplification) between introduced noise and other image degradations (\eg, blur).

\subsection{Blur}\label{section:blur}

Blur ($\mathcal{B}_\mathbf{k}$) is also one of the most common degradations appearing in image capturing \eg, camera shake in handheld photography, motion blur, and defocus blur~\cite{yuan2007imagedeblur, abuolaim2022improvingdeblur, hosseini2019convolutionalblur}. Capturing aligned blurry-clean pairs in real scenarios is extremely difficult, for this reason, the most popular deblurring datasets are synthetic. 
Most approaches adopt a uniform blur by convolving the image with iso/anisotropic Gaussian kernels~\cite{kai2021bsrgan}. Xu~\etal~\cite{xu2019rawsr, xu2020exploiting} implemented defocus blur as a disk kernel, and a modest motion blur~\cite{schuler2013machineblur}.

We create our blur degradation by convolving the image with a \textbf{diverse pool of kernels}: classical isotropic and anisotropic gaussian blur kernels~\cite{kai2021bsrgan, levin2009understanding}, real estimated motion blur kernels~\cite{pan2016blind, xu2017motionblur, ren2020selfdeblur}, and real estimated PSFs (point-spread-function) from~\cite{zhang2020usrnet, zhu2012deconvolvingblur, xu2010twoblur}.

This task is fundamental since the current trend in reducing size and weight observed for the most popular DSLR or mirrorless cameras, and for the smartphone cameras, makes it challenging to implement sensor stabilization systems handling extreme levels of sensor instability during point-and-shoot acquisition. 

\subsection{Exposure}

Exposure levels (EE) can introduce undesirable artifacts in photographic images. Both underexposure and overexposure can result in signal loss or increased noise in the imagery~\cite{chen2018learning, hasinoff2016burst, chang2021low}. While our inclusion of this degradation isn't intended to address HDR issues~\cite{hasinoff2016burst}, it does allow the model to acquire exposure correction skills. Additionally, the interplay between simulated exposure levels and noise augments the complexity of the degradation environment.
We adopt a low exposure model inspired by the work of Punnappurath \etal~\cite{punnappurath2022day}, employing a linear ISO scaling technique to generate the resultant darkened image. It is worth noting that our real-world testing dataset comprises images with exposure-related challenges, along with their respective, carefully-selected ground truth.

\subsection{Downsampling}
\label{section:downsampling}
One of the most important differences between cameras is the size of the sensor~\cite{karaimer2016software, ignatov2020replacing, delbracio2021mobile}.
Photographers can use telephoto lenses to obtain HR images, but the resolution of the scene is limited by the size of the sensor. Thus, it is more common to capture a wide-range scene as a LR image using a wide-angle lens, and then apply SISR~\cite{xu2019rawsr, xu2020exploiting}. Considering this, it is desirable to be able to upscale RAW images~\cite{xu2019rawsr, xu2020exploiting}. However, this is not a trivial task -as for RGB images- since most traditional methods do not preserve the Bayer pattern (RGGB).

Since obtaining real LR-HR pairs is extremely expensive and time-consuming, we follow~\cite{xu2019rawsr, kai2021bsrgan} to synthesize LR RAW images from the assumed HR real captures (\eg 60MP RAWs). We downsample the high-quality RAW images so that each pixel could have its ground truth red, green and blue values. We denote this operation as $\downarrow_s$.

\begin{figure*}[!ht]
    \centering
    \setlength\tabcolsep{1.0pt}
    \begin{tabular}{ccc}
         \includegraphics[width=0.32\linewidth]{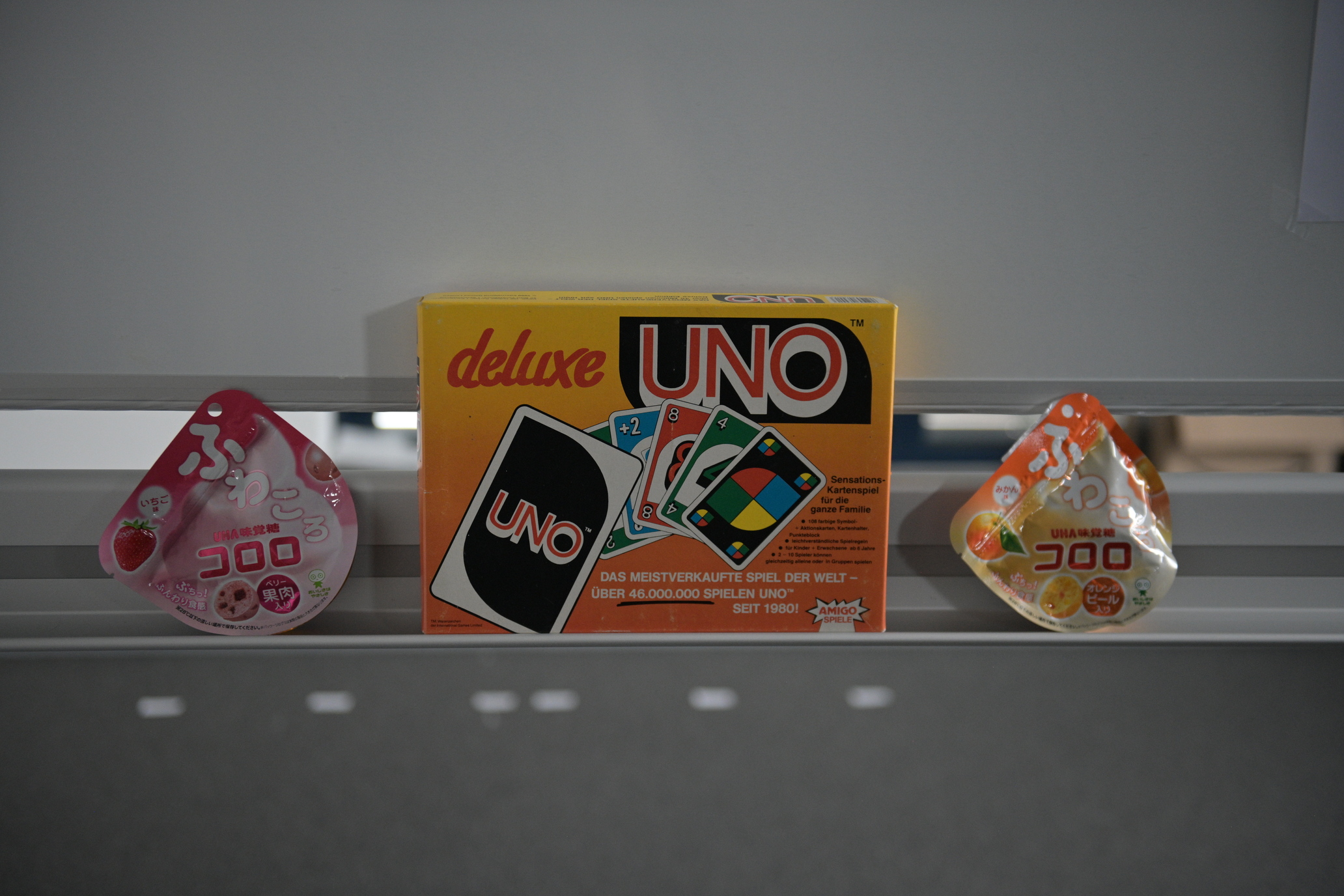} &
         \includegraphics[width=0.32\linewidth]{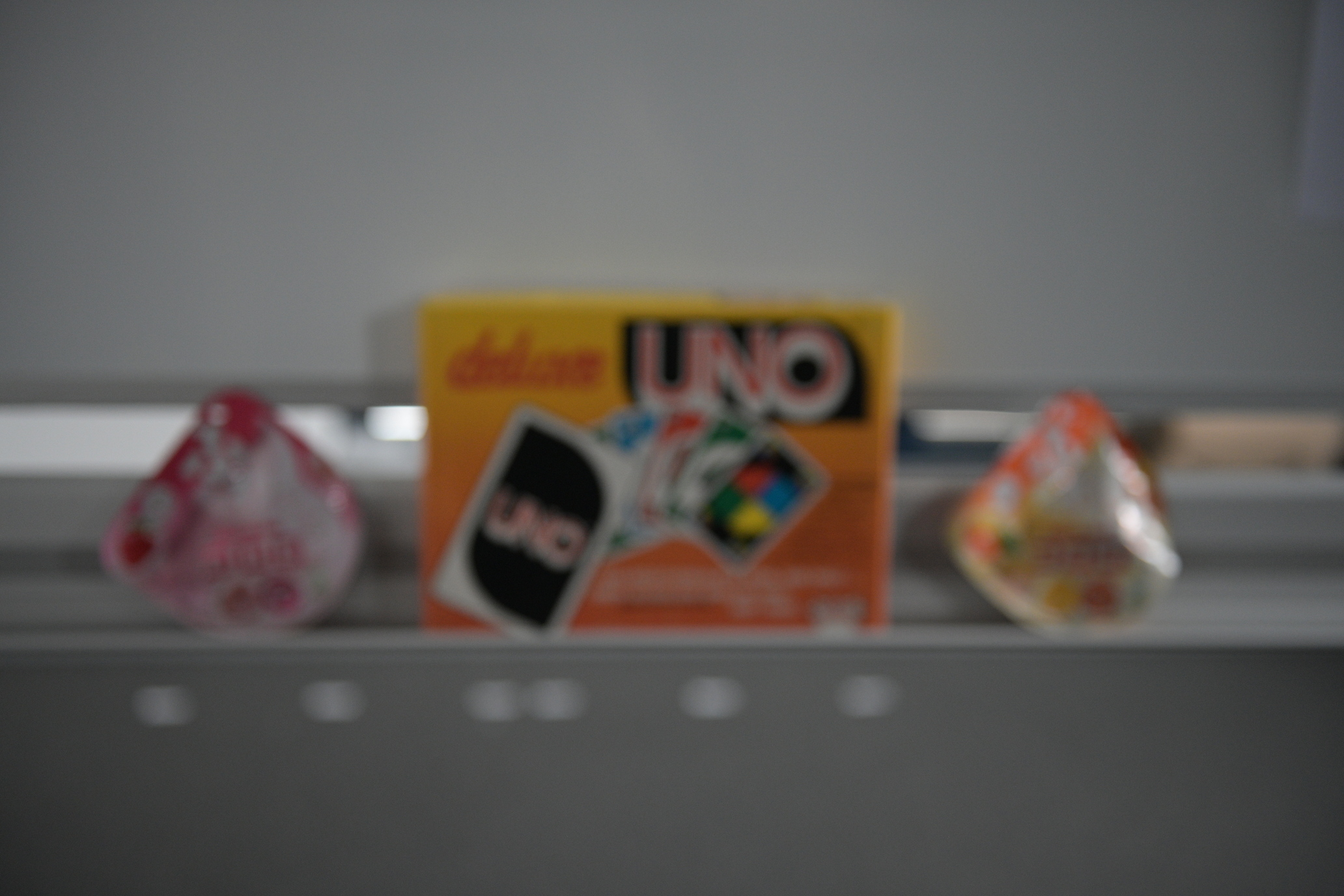} &
         \includegraphics[width=0.32\linewidth]{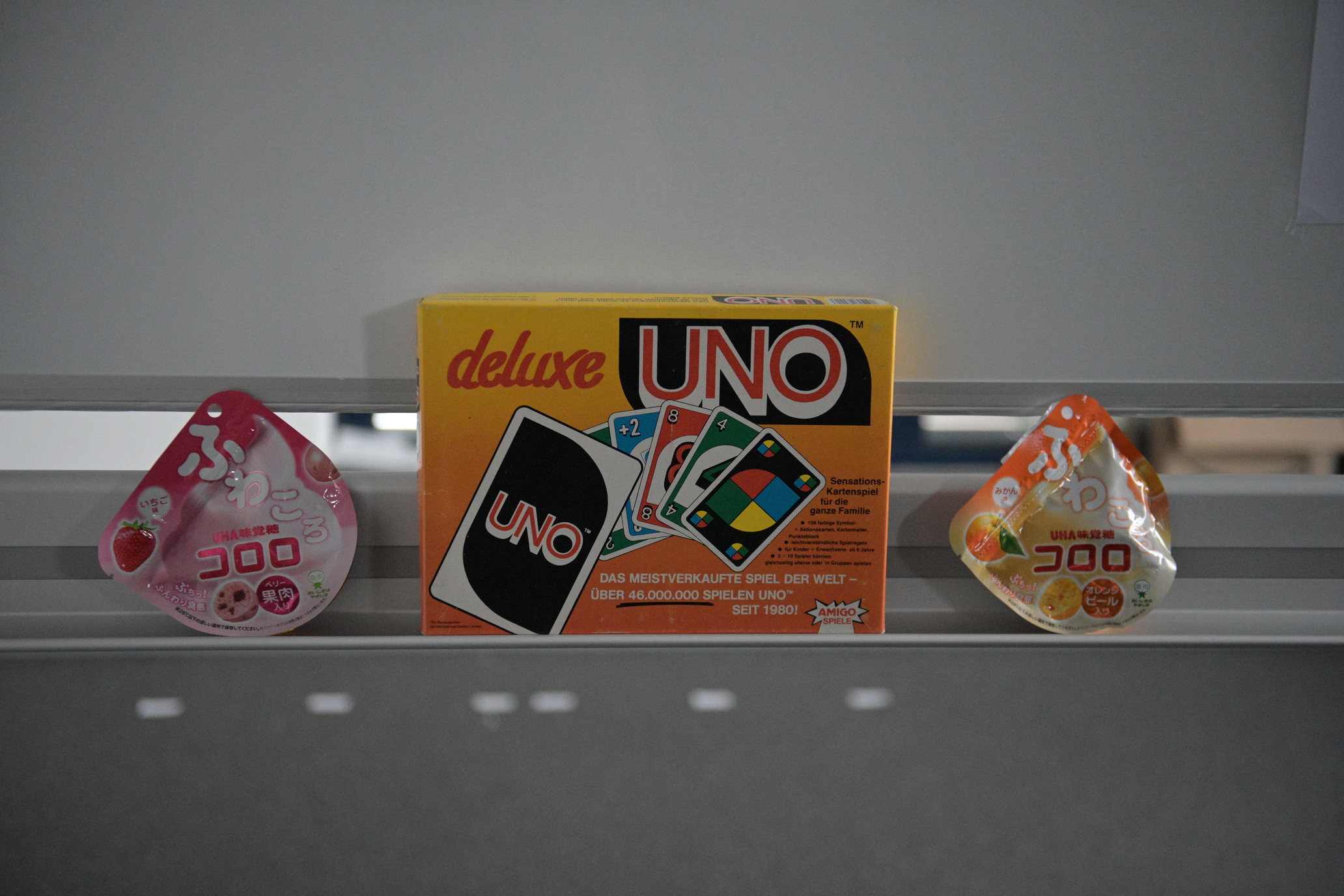} \tabularnewline
         Ground Truth   & Defocus Blur & High ISO Noise
         \tabularnewline
    \end{tabular}
    \caption{Real world samples representing usual RAW image defects. Best in the electronic format. Images captured manually with our Nikon Z7 on a tripod. Illumination and stabilization are controlled to ensure proper alignment.}
    \label{fig:real-world-samples}
\end{figure*}

\subsection{Complete RAW Degradation Model}
\label{ssc:random_shuffle}

The degradation pipeline covers most of the factors affecting the quality of RAW images. One important aspect is the noise sampling from multiple noise profiles, as well as the PSF sampling (and even application of two PSFs). The order of the degradations is defined by the optics theory~\cite{elad1997restoration} \ie first we apply the kernel ($\mathbf{k}$) to the RAW signal, then we simulate a small sensor with downsampling and noise injection -that can be influenced by the exposure- and later we apply the other degradations inclusing additional motion blur and downsampling.

We define four different degradation levels in our experiments: \RNum{1}) classical image restoration~\cite{liang2021swinir, elad1997restoration}. Only blur and noise are considered. \RNum{2}) SISR degradation model (see Eq.~\ref{eq:sisr_degradation}) \RNum{3}) SISR considering additional exposure degradations \RNum{4}) Finally we apply our complete realistic degradation pipeline, including multiple noise profiles, the application of multiple kernels, exposure, and downsampling. We provide more details about the pipeline in the appendix.

\section{Experiments}

We conduct several experiments to prove the benefits of our approach for RAW image restoration and SISR. We propose BSRAW as an efficient baseline method for this task. Due to the pages limit, we provide additional implementation details, dataset information, and results in the supplementary material.

\subsection{BSRAW Dataset}
\label{sec:datasets}

Our goal is to to solve blind super-resolution for RAW images, and apart from a realistic degradation prior (Sec.~\ref{sec:method}), an image prior also contributes to the success of the deep model~\cite{kai2021bsrgan}. 
Our sources of real RAW data and our train/test split are summarized in Table~\ref{tab:datasets}. The quality of the DSLR RAW images is substantially better than in smartphones due to the larger optics and sensor, this implies that the RAW images show less noise and minimal optical aberrations.

\begin{table}[t]
    \centering
    \resizebox{\columnwidth}{!}{
    \begin{tabular}{l l c c c c}
         \toprule
         \textbf{Device} & \textbf{Sensor} & \textbf{\# Train}  & \textbf{\# Test} & \textbf{Res.} \\
         \midrule
         DSLRs~\cite{fivek, xu2019rawsr}    & Diverse & 1000 & 150 & 10MP \\
         \midrule
         \rowcolor{Gray} Sony $\alpha$7R4          & IMX551     & 90    & 10 & 2MP \\
         \rowcolor{Gray} Nikon Z7                  & IMX309BQJ  & 90    & 10 & 2MP \\
         \bottomrule
    \end{tabular}}
    \caption{Dataset split. We use diverse DSLR images from the MIT5K as previous works~\cite{fivek, xu2019rawsr} (Canon EOS 5D and Nikon D700). The sensors highlighted in gray correspond to our new dataset of real-world captures. The train/test split ensures no overlap and proper evaluation.
    \vspace{-2mm}
    }
    \label{tab:datasets}
\end{table}

Following previous work~\cite{xing2021invertible, xu2019rawsr, xu2020exploiting}, we use images from the Adobe MIT5K dataset~\cite{fivek}, which includes two DSLR cameras. 
The DSLR images are manually filtered to ensure diversity and natural properties (\ie remove extremely dark or overexposed images), we also remove the blurry images (\ie we only consider all-in-focus images).

The \textbf{pre-processing} is as follows: (i) we normalize all RAW images depending on their black level and bit-depth. (ii) we convert (``pack") the images into the well-known RGGB Bayer pattern (4-channels), which allows to apply the transformations and degradations without damaging the original color pattern information~\cite{liu2019learningrawaug}. (iii) Following previous work on RAW processing and learned ISPs~\cite{xu2019rawsr, ignatov2021learnednpu}, we train using image patches of dimension $248\times248\times4$. 

\vspace{-2mm}
\paragraph{Synthetic Dataset} We apply our complete degradation pipeline (see Sec.~\ref{ssc:random_shuffle}) to the training images to generate aligned degraded-clean pairs. The synthetic test dataset is generated by applying our degradation pipeline, at different levels, to the corresponding test images -- see Table~\ref{tab:datasets}. 

\vspace{-3mm}
\paragraph{Real-World Captures}
We believe the models trained using our pipeline have notable generalization capabilities (also concluded in~\cite{kai2021bsrgan, xu2019rawsr}), therefore, we collect a novel dataset with challenging real scenes. As indicated in Table~\ref{tab:datasets}, we capture images using two mirrorless full-frame cameras: Sony $\alpha$7R4 (60MP) and Nikon Z7 (45.7MP). In Figure~\ref{fig:real-world-samples} we show samples from our data capturing setup.

This is a paired dataset consisting of variations over 100 real-world indoor and outdoor scenes. All the devices used for data capturing allow for manual operation of the camera ISP parameters and optics, and are fixed on a tripod. For each ``clean" reference image, we capture several variations, where each variation represents a degradation with a different degree of intensity. The scene variants represented there are degraded in terms of defocusing blur, exposure time, high ISO noise, and motion blur. Since interaction with the camera device was needed during the acquisition, we perform a post-processing alignment (similar to~\cite{bhat2021deep}) to further improve the quality and usability of the data.  

\begin{figure}[t]
     \centering
     \begin{subfigure}[b]{0.49\textwidth}
         \centering
         \includegraphics[trim={0.5cm 0 0 0},clip,width=\textwidth]{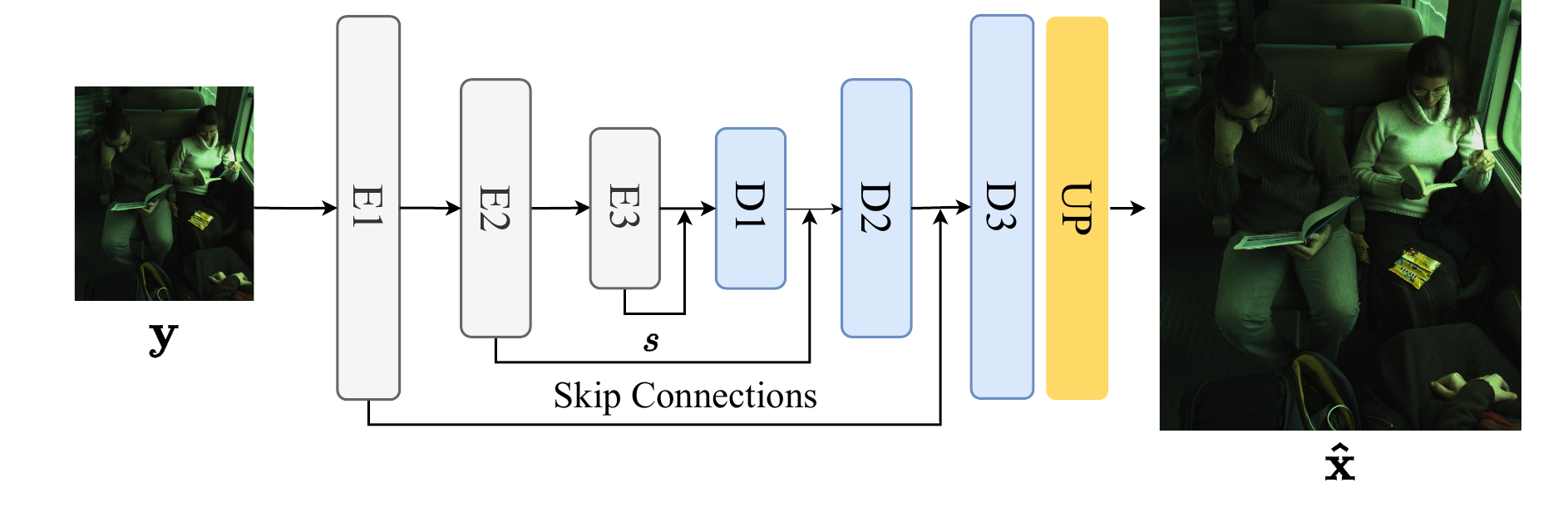}
     \end{subfigure}
     \hfill
     \begin{subfigure}[b]{0.49\textwidth}
         \centering
         \includegraphics[width=\textwidth]{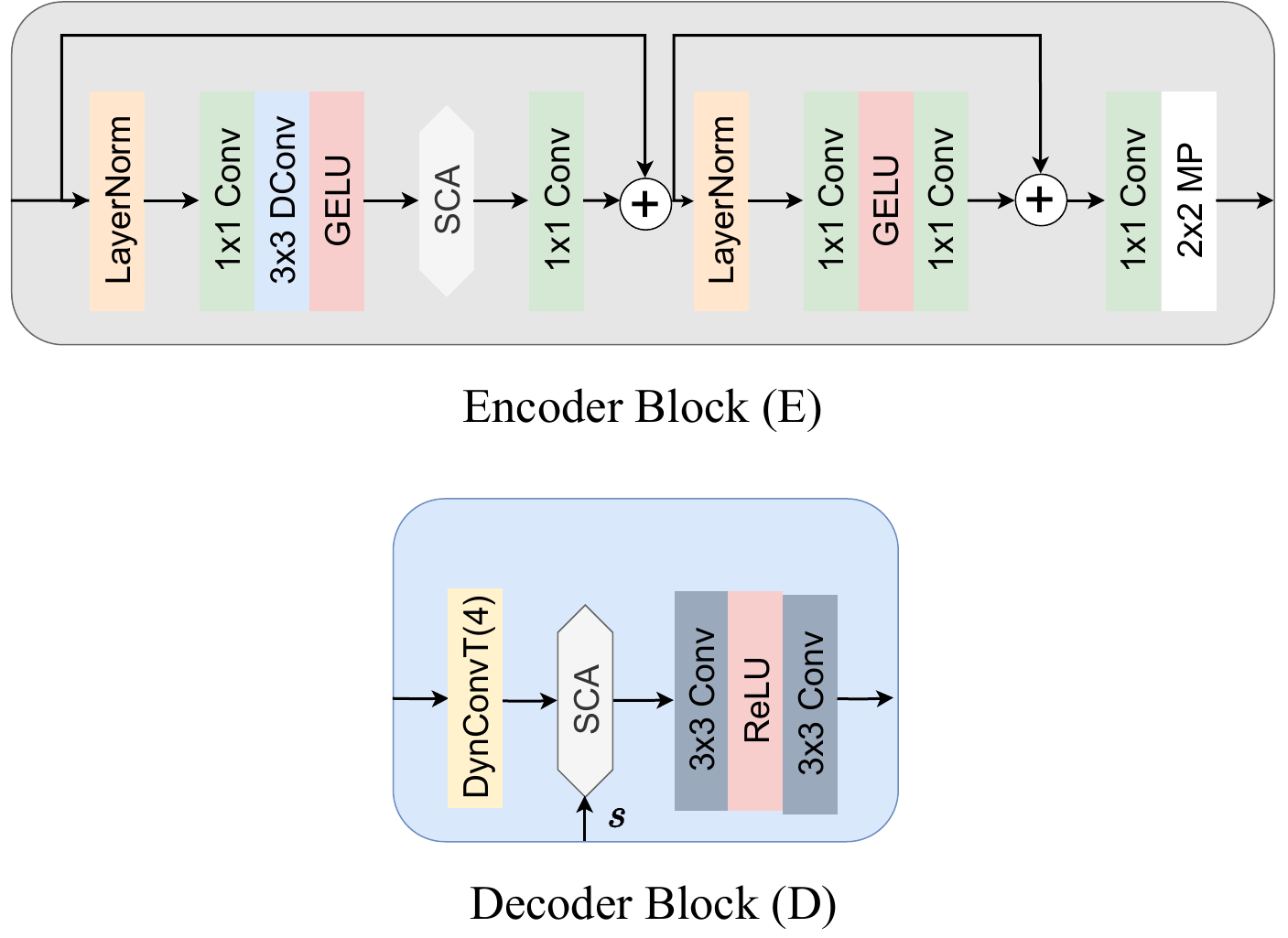}
     \end{subfigure}
     
\caption{Architecture of the proposed BSRAW inspired in NAFNet~\cite{chen2022simple}. We simplify the encoder blocks and use Dynamic convolutions to process images more efficiently.}
\label{fig:model}
\end{figure}

\subsection{Experimental Results} 

We propose BSRAW as an efficient baseline, able to process 4K resolution images in standard GPUs (120 ms for a 4K image on a Nvidia 3090Ti 24GB GPU), without patching or tiling strategies. The architecture is illustrated in Figure~\ref{fig:model}. BSRAW combines NAFNet~\cite{chen2022simple} baseline blocks with dynamic convolutions and simplified stereo channel attention in the decoder part of the model. In comparison to NAFNet~\cite{chen2022simple} -already quite efficient- our method has $10\times$ less parameters and $\approx20\%$ less MACs.  We upscale the images using an adapted pixel-shuffle head.

\paragraph{SISR results} We compare with previous method RAWSR~\cite{xu2019rawsr} for SISR. Our realistic degradation pipeline helps our efficient model BSRAW to generalize better on complex degradations. We upscale the images using BSRAW, and later process them using a basic ISP~\cite{karaimer2016software} \eg Dcraw as~\cite{xu2019rawsr} to obtain the resultant RGB. Note that the training dataset derived from MIT5K~\cite{fivek} from Xu~\etal RAWSR~\cite{xu2019rawsr} is not publicly available. Therefore we train using our own curated DSLR dataset (see Table~\ref{tab:datasets}); this, together with BSRAW being trained using our degradation pipeline, allow us to achieve notably better results.  We also compare with TENet~\cite{qian2019trinity} (based on RDN~\cite{zhang2018residualdense}) pipeline for ``Denoising (DN) $\rightarrow$ SR" for RAWs. 

We present the results in Table~\ref{tab:rawsr}. We distinguish three variants of our model, the \textbf{ablation study} is as follows:

(i) BSRAW~\RNum{2} is trained using a simple SISR degradation pipeline, similar to~\cite{xu2019rawsr} -our level \RNum{2}-.
(ii) BSRAW~\RNum{4} is trained using our complete degradation pipeline. We can appreciate how this helps to improve results notably.
Finally, (iii) BSRAW~$\dagger$ extends the previous variant, and we include training data from our dataset. This leads to the best results in terms of fidelity metrics.

In Table~\ref{tab:expx2} we show the results from our model on the synthetic dataset at each degradation level.

\begin{table}[t]
    \centering
    \begin{tabular}{l c c c}
        \toprule
        \textbf{Methods} & \textbf{PSNR~$\uparrow$}  & \textbf{SSIM~$\uparrow$} & \textbf{Domain} \\
        \midrule
        DeepISP & 21.71 & 0.7323 & RAW \\
        RDN~\cite{zhang2018residualdense} & 29.93 & 0.7804 & RGB \\
        TENet~\cite{qian2019trinity}      & 30.20 & 0.800  & RAW \\
        BSRGAN~\cite{kai2021bsrgan}       & 30.01 & 0.794  & RGB \\
        RAWSR~\cite{xu2019rawsr}          & \underline{30.79} & \underline{0.804}  & RAW+RGB \\
        \midrule
        \rowcolor{Gray} BSRAW~\RNum{2} \emph{(Ours)}             & 31.02 & 0.812  & RAW \\
        \rowcolor{Gray} BSRAW~\RNum{4} \emph{(Ours)}             & \textbf{31.40} & \textbf{0.843}  & RAW \\
        \rowcolor{Gray} BSRAW~$\dagger$ \emph{(Ours)}            & 31.86 & 0.860  & RAW \\
        \bottomrule
    \end{tabular}
    \caption{Comparison for RAW SISR $\times2$. We use the DSLR testset from~\cite{xu2019rawsr}, which consists on 150 images from MIT5K~\cite{fivek} degraded with a simple SISR model~\cite{xu2019rawsr}. Our model BSRAW achieves \emph{state-of-the-art} results thanks to our degradation pipeline and our dataset. Metrics calculated on RGBs. We process the upscaled RAWs using a fixed ISP and metadata~\cite{xu2019rawsr} \eg Dcraw.
    }
    \label{tab:rawsr}
\end{table}

\begin{figure}[!ht]
    \centering
    \setlength{\tabcolsep}{0pt}
    \begin{tabular}{cc}
    \includegraphics[width=0.45\linewidth]{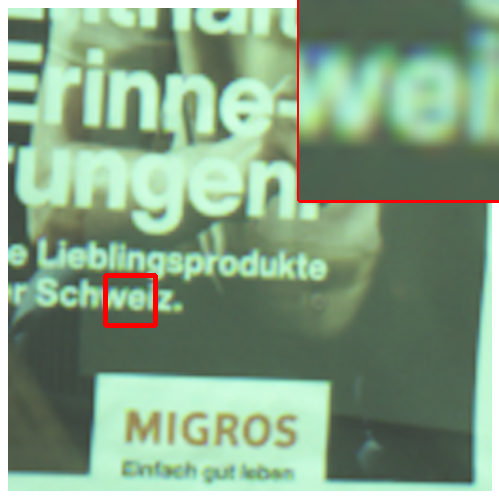} &
    \includegraphics[width=0.45\linewidth]{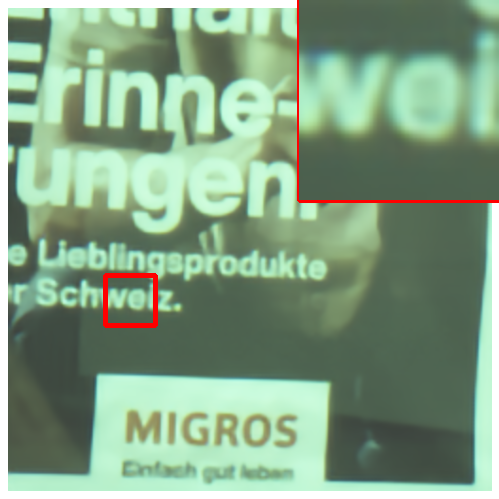} \\ 
    \includegraphics[width=0.45\linewidth]{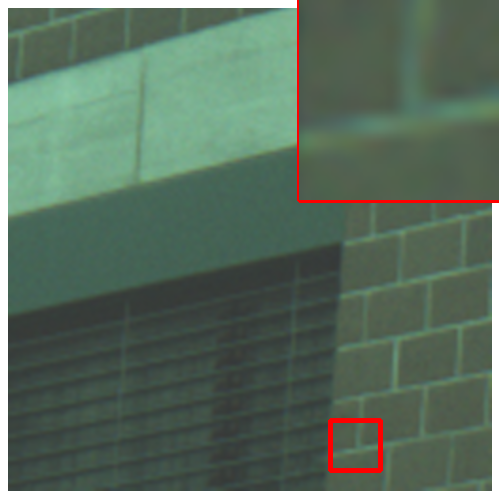} &
    \includegraphics[width=0.45\linewidth]{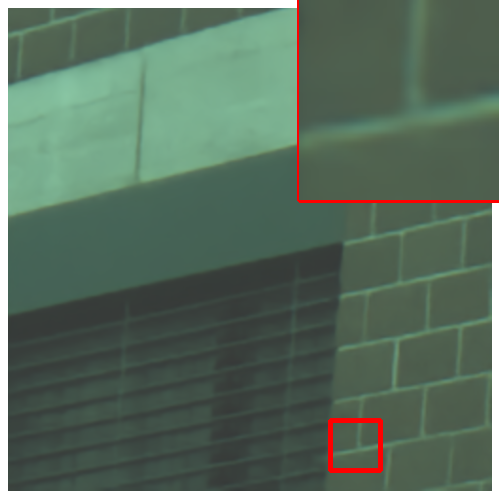} \\
    \includegraphics[width=0.45\linewidth]{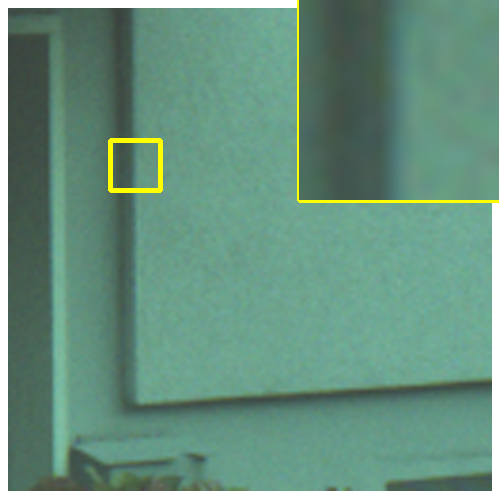} &
    \includegraphics[width=0.45\linewidth]{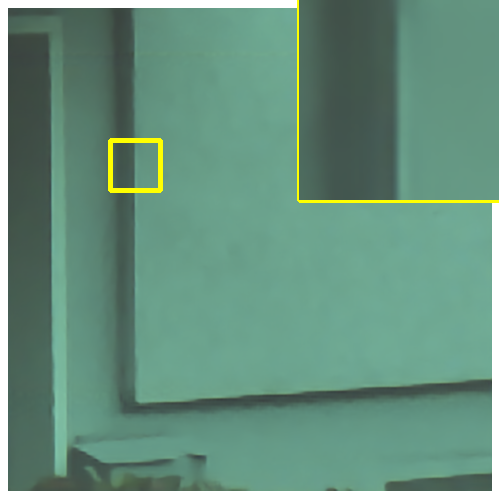} \\
    \includegraphics[width=0.45\linewidth]{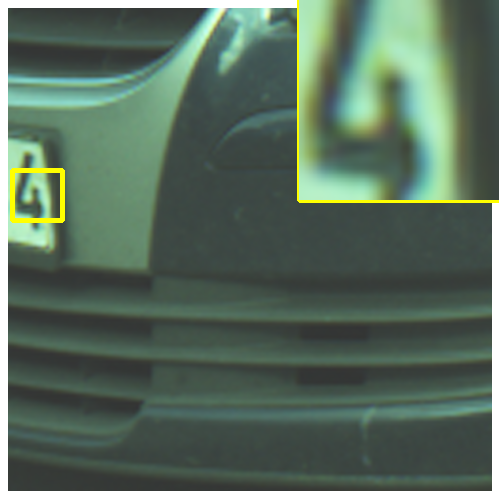} &
    \includegraphics[width=0.45\linewidth]{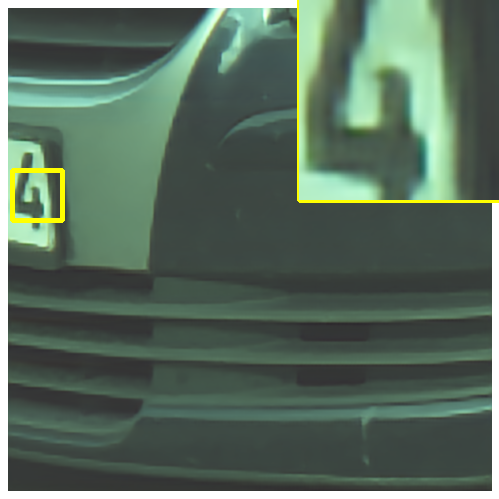} \\ 
    Interpolation & BSRAW (Ours) \tabularnewline
    \end{tabular}
    \caption{Our BSRAW model, thanks to the proposed degradation pipeline, can perform \textcolor{red}{2$\times$ (red)} and \textcolor{orange}{4$\times$ (yellow)} SISR and restore RAW images in the wild. We use images from a smartphone Sony IMX586 Quad Bayer sensor~\cite{ignatov2021learnednpu}, which was not used for training (neither similar sensors). This proves certain generalization capabilities, even on unseen sensors.}
    \label{fig:sony}
\end{figure}

\begin{figure*}[!ht]
    \centering
    \begin{tabular}{c}
         \includegraphics[width=\linewidth]{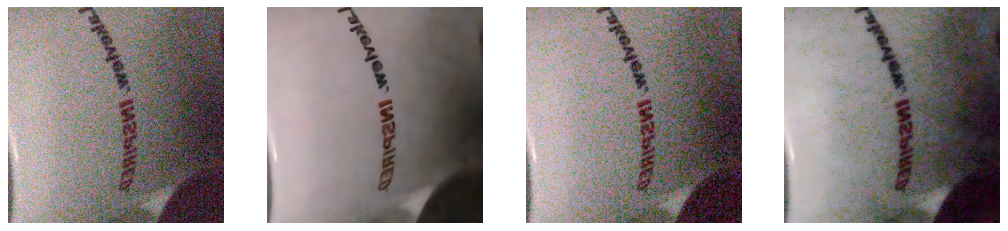} \\
         \includegraphics[width=\linewidth]{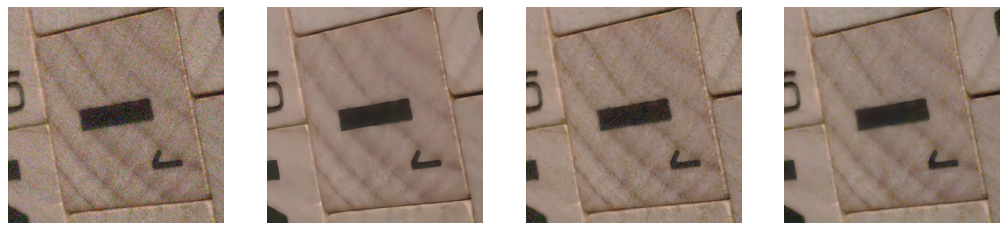} \\
         \includegraphics[width=\linewidth]{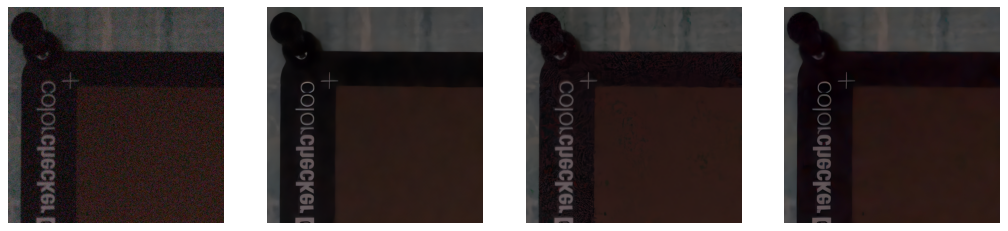} \\
         \hspace{10pt} Input image \hspace{75pt} BSRAW (Ours) \hspace{60pt} BSRGAN~\cite{kai2021bsrgan} \hspace{60pt} Real-ESRGAN~\cite{wang2021realESRGAN,cai2019realsr}
    \end{tabular}
    \caption{From left to right: sRGB rendered using the input RAW and a simple ISP~\cite{abdelhamed2018high}, our restored RAW (we apply the ISP after restoring and upsampling), BSRGAN~\cite{kai2021bsrgan} and Real-ESRGAN~\cite{wang2021realESRGAN,cai2019realsr} restored images. The approaches are illustrated in Figure~\ref{fig:main-wild}. We believe both blind single image super-resolution (SISR) approaches might be complementary and represent future research.}
    \label{fig:rgb_samples}
\end{figure*}

\paragraph{Effectiveness on Real Images} As we show in Figures~\ref{fig:rgb_samples} and Table~\ref{tab:rawsr} for RAW restoration and SISR, our model can generalize and provide reasonable results on real scenes. Figures~\ref{fig:sony} and~\ref{fig:rgb_samples} qualitative results show that BSRAW is able to produce noise-free sharp images with proper textures and color distributions, even using RAW images from unseen sensors such as Sony Quad sensors or SIDD phones~\cite{abdelhamed2018high}.

\vspace{2mm}
\emph{Super-Resolution, RAW or RGB?} We also compare the following two strategies for image processing:\\

\textbf{(I)} SwinIR~\cite{liang2021swinir} and BSRGAN~\cite{kai2021bsrgan}, general-purpose models trained for real-world blind SISR on RGBs. Their input is a degraded RGB image obtained after applying an ISP to the original degraded RAW image. We use images from the SIDD dataset~\cite{abdelhamed2018high} with noise and low exposure.\\

\textbf{(II)} Our approach. BSRAW takes directly the degraded RAW image, applies restoration and upsampling, then we apply the ISP with known parameters and operations.\\

\noindent The ISP~\cite{karaimer2016software} is fixed for a fair comparison. The images are real-world captures from ``unknown" sensors to the model.

In Figure~\ref{fig:rgb_samples} we compare these two possible variants of super-resolution. The quantitative results form Table~\ref{tab:rawsr}, and qualitative results (Figure~\ref{fig:rgb_samples}) show the advantages and potential of our approach. We provide further analysis and samples in the supplementary material.

\vspace{4mm}

\subsection{Implementation Details} 

The method was implemented in PyTorch. We use the basic $\mathcal{L}_1$ loss between the ground-truth and the restored image. We set the minibach size to 32. We use Adam~\cite{kingma2014adam} optimizer with standard hyper-parameters. The initial learning rate is set to $1e^{-4}$, we use a linear LR decay scheduler, until a minimum of $1e^{-5}$. We perform experiments using multiple NVIDIA RTX 3090Ti and 4090Ti.

The degradation pipeline can be integrated into the dataloader as an augmentation. All the degradations are applied on 4-channel RAW images (RGGB).

Note that the training dataset derived from MIT5K~\cite{fivek} from Xu~\etal RAWSR~\cite{xu2019rawsr} is not publicly available, and therefore cannot be used for comparison. We cannot perform a qualitative comparison because RAWSR weights and results are not publicly available \ie the material at \footnote{\url{https://github.com/xuxy09/RAWSR}} is obsoleted and links do not work.

In the \emph{RAW domain}, we can compare the HR RAW $(1460 \times 2192 \times 4)$ and the LR RAW $(365 \times 548 \times 4)$ --- which correspond to RGBs of resolution $(2920 \times 4384 \times 3)$ and $(730 \times 1096 \times 3)$ respectively. Upscaling the LR RAW using Bicubic interpolation implies a reconstruction of 27.13dB (and 0.81 SSIM) \emph{w.r.t} the HR RAW. 

\subsection{Limitations}

Downsampling RAW images is not trivial, this operation can cause color artifacts besides blurriness, and irreversible structure loss. Also, the MIT5K~\cite{fivek} images were capture with ``old" cameras, for instance, 12MP DSLRs, in comparison with our 60MP DSLM cameras. We need further analysis using modern cameras, specially smartphones.

\section{Conclusion}
\label{sec:conclusion}

In this paper we discuss a controllable degradation pipeline to synthesize realistic degraded RAW images for training deep blind super-resolution models.
We have curated a dataset with different DSLR cameras. As part of this effort, we provide a new dataset of diverse real scenes captured using DSLM cameras.
Our experiments demonstrate that models trained with our degradation pipeline can restore real-world degraded RAWs. This represents a powerful alternative in image signal processing, and can be of benefit to other low-level downstream tasks.

\paragraph{Future Work}
We plan to extend this analysis to smartphone images, where the ISP is more challenging, and the possible degradations in the RAW images might be more complex due to the limited optics \ie more noise and blur.

\paragraph{Acknowledgments} This work was partly supported by the The Alexander von Humboldt Foundation (AvH).

\begin{table*}[!ht]
    \centering
    \resizebox{\linewidth}{!}{
    \begin{tabular}{c c c c c || c c c c }
        \toprule
        \rowcolor{Gray} & \multicolumn{4}{c||}{\bf Super-Resolution $\times2$} & \multicolumn{4}{c}{\bf Super-Resolution $\times4$} \\
        \midrule
        Degradation & \multicolumn{2}{c}{BSRAW-light} & \multicolumn{2}{c||}{BSRAW} & \multicolumn{2}{c
        }{BSRAW-light} & \multicolumn{2}{c}{BSRAW} \\
        & PSNR$\uparrow$ & SSIM$\uparrow$ &  PSNR$\uparrow$ & SSIM$\uparrow$ & PSNR$\uparrow$ & SSIM$\uparrow$& PSNR$\uparrow$ & SSIM$\uparrow$ \\
        \midrule
        \RNum{1} & 42.57 & 0.949 & 43.30 & 0.952  & 40.08 & 0.929 & 40.51 & 0.937\\
        \RNum{2} & 43.26 & 0.961 & 43.98 & 0.965  & 40.42 & 0.941 & 41.28 & 0.949\\
        \RNum{3} & 43.53 & 0.962 & 43.96 & 0.966  & 40.51 & 0.939 & 41.34 & 0.951\\
        \RNum{4} & 43.78 & 0.964 & 44.25 & 0.971  & 39.71 & 0.927 & 41.45 & 0.957 \\
        \bottomrule
    \end{tabular}
    }
    \caption{Ablation study for BSRAW on different degradation levels using our \emph{synthetic MIT5K testset}. The metrics are calculated in the RGB domain, thus, the ISP setting was fixed for all the compared configurations, to avoid possible biases, we use a simple ISP with the canonical steps (demosaicing, white balance, color correction, gamma and tone mapping correction.)~\cite{karaimer2016software}. BSRAW-light is a variant with less encoder blocks and channels, described in the supplementary material.
    \vspace{8mm}}
    \label{tab:expx2}
\end{table*}

{\small
\bibliographystyle{ieee_fullname}
\bibliography{egbib}
}

\end{document}